\documentclass[twocolumn,final,twoside,journal,a4paper]{IEEEtran}

\usepackage[usenames,dvipsnames]{xcolor}
\usepackage{amsmath,amsthm,graphicx,cite}
\usepackage{epsfig,amsfonts}
\usepackage{graphicx,cite,amssymb,amsmath}
\usepackage{color}
\usepackage{enumitem}

\usepackage[nolist]{acronym}
\usepackage{psfrag}
\usepackage{perso}
\usepackage{xcolor}
\usepackage{blindtext}
\usepackage[keeplastbox,ancient]{flushend}
\usepackage{mathtools}
\mathtoolsset{showonlyrefs}
\newcommand{\nfiles}{n}
\newcommand{\npcached}[1]{w_{#1}}

\newcommand{\npacket}{k}
\newcommand{\codedpacket}{c}

\newcommand{\indexi}{i}
\newcommand{\coeff}{g}

\newcommand{\radius}[1]{r_{#1}}

\newcommand{\dcenter}{d}
\newcommand{\memory}{M}

\newcommand{\sEvent}{S}

\newcommand{\lowerb}[1]{l{#1}}
\newcommand{\upperb}[1]{u{#1}}

\newcommand{\Pro}{P}

\newcommand{\library}{\mathcal{F}}
\newcommand{\libraryf}[1]{\mathsf{f}_{#1}}

\newcommand{\hubs}{h}

\newcommand{\probHub}[1]{\gamma_{#1}}
\newcommand{\probFile}[1]{\theta_{#1}}

\newcommand{\probSuc}[1]{P_s{#1}}
\newcommand{\probFail}[1]{P_f{#1}}

\newcommand{\isymb}{u}
\newcommand{\isymbv}{\mathbf{\isymb}}

\newcommand{\rosymb}{y}
\newcommand{\rosymbv}{\mathbf{\rosymb}}
\newcommand{\G}{\mathbf{G}}
\newcommand{\Grx}{\tilde{ \mathbf{G}}}
\newcommand{\Expd} [1]{\mathbb{E}[\Delta_{#1}] }

\begin{document}

\begin{acronym}

\acro{LRFC}{linear random fountain code} 
\acro{GEO}{geostationary Earth orbit}
\acro{ILP}{Integer Linear Programming}
\acro{LP}{Linear Programming}
\acro{MDS}{maximum distance separable}
\acro{HAPS}{high altitude platform station}

\end{acronym}

%%%%%%%%%%%%%%%%%%%%%%%%%%%%%%%%%%%%%%%%%%%%%%%%%%%%%%%%%%%%%%%%%%%%%%%%%%%%%%%%%%%%%%%%%%%%%%%%%%%%
%%%%%%%%%%%%%%%%%%%%%%%%
\title{Caching at the Edge with Fountain  Codes}

\author{
    \IEEEauthorblockN{Estefan\'ia Recayte, Francisco L\'azaro, Gianluigi Liva\\
    \IEEEauthorblockA{\IEEEauthorrefmark{1}Institute of Communications and Navigation of DLR (German Aerospace Center),
    \\Wessling, Germany. Email:  \{Estefania.Recayte, Francisco.LazaroBlasco,Gianluigi.Liva\}@dlr.de}\\
}
\thanks{This work has been accepted for publication at IEEE ASMS 2018.}
\thanks{\copyright 2018 IEEE. Personal use of this material is permitted. Permission
from IEEE must be obtained for all other uses, in any current or future media, including
reprinting /republishing this material for advertising or promotional purposes, creating new
collective works, for resale or redistribution to servers or lists, or reuse of any copyrighted
component of this work in other works.}

}
\maketitle

%\maketitle
%%%%%%%%%%%%%%%%%%%%%%%%%%%%%%%%%%%%%%%%%%%%%%%%%%%%%%%%%%%%%%%%%%%%%%%%%%%%%%%%%%%%%%%%%%%%%%%%%%%%
%%%%%
%\tableofcontents

%\thispagestyle{empty}

%%%%%%%%%%%%%%%%%%%%%%%%%%%%%%%%%%%%%%%%%%%%%%%%%%%%%%%%%%%%%%%%%%%%%%%%%%%%%%%%%%%%%%%%%%%%%%%%%%%%
%%%%%%%%%%%%%%%%%%%%%%%%%
\thispagestyle{empty} \pagestyle{empty}

\begin{abstract}
We address the use of \acl{LRFC} caching schemes in a heterogeneous satellite network. We consider a system composed of multiple hubs and a geostationary Earth orbit satellite. Coded content is memorized in hubs' caches in order to serve immediately the user requests and reduce the usage of the satellite backhaul link.  We derive the analytical expression of the average backhaul rate, as well as a tight upper bound to it with a simple expression. Furthermore, we derive the optimal caching strategy which minimizes the average backhaul rate and compare the performance of the linear random fountain code scheme to that of a scheme using \acl{MDS} codes. Our simulation results indicate that the performance obtained using fountain codes is similar to that of \acl{MDS} codes.
\end{abstract}

%%%%%%%%%%%%%%%%%%%%%%%%%%%%%%%%%%%%%%%%%%%%%%%%%%%%%%%%%%%%%%%%%%%%%%%%%%%%%%%%%%%%%%%%%%%%%%%%%%%%%%%%%%

%\begin{keywords}
%\end{keywords}

%%%%%%%%%%%%%%%%%%%%%%%%%%%%%%%%%%%%%%%%%%%%%%%%%%%%%%%%%%%%%%%%%%%%%%%%%%%%%%%%%%%%%%%%%%%%%%%%%%%%%%%%%%

\section{Introduction}\label{sec:Intro}

Cache-aided delivery protocols represent a  promising solution to counteract the dramatic increase in demand for multimedia content in wireless networks.  Caching techniques have been widely studied in literature with the aim of reducing the backhaul congestion,  the energy consumption and the latency. In cache-enabled networks content is pre-fetched close to the user during network off-peak periods in order to directly serve the users when the network is congested.
In \cite{Maddah2015}  Maddah-Ali \emph{et al.}    aim at reducing the transmission rate in a network where each user has an individual cache memory.  In that work,  the idea of \emph{coded caching} is introduced, so that the cache memory not only  provides direct local access to the content but also  generates coded multicasting opportunities among users requesting different files.

In \cite{bioglio:Globcom2015} and  \cite{Liao:2017}, \ac{MDS} codes are proposed for minimizing the use of the backhaul link during the delivery phase in networks with caches at the transmitter side only.  In \cite{ozfatura2018} a delayed offloading scheme based on \ac{MDS} codes is proposed to spare backhaul link resources in a network with mobile users. Caching schemes leveraging on \ac{MDS} codes have also been proposed  for device to device communication in order to reduce the latency  \cite{piemontese2016}.

\ac{MDS} codes, such as Reed-Solomon codes, are optimal in the sense that they achieve the Singleton bound. The drawback of such codes is their limited flexibility in the choice of the code parameters (e.g. the block length) once the finite field is fixed, and the fact that the rate of the code is set before the encoding takes place. Unlike \ac{MDS} codes, fountain codes \cite{MacKay05:fountain}  are \emph{rateless}, i.e., their rate can be adapted on-the-fly. This has the advantage of adding flexibility to the network, allowing a dynamic resource management.
%Furthermore, the loss in efficiency of fountain codes with respect to \ac{MDS} codes can be mitigated by using codes over extended Galois fields \cite{lazaro:Allerton2015}.
%As pointed out in \cite{bioglio:Globcom2015}, fountain codes are less efficient than \ac{MDS} codes.

Extensive studies regarding caching for terrestrial
applications can be found in literature while limited work is available in the context of heterogeneous satellite networks  \cite{DeCola:icn, Wu:Twolayer, Liu:LeoGame, Kalantari:SatTerresrial}.   In \cite{Kalantari:SatTerresrial} a  off-line caching approach over a hybrid satellite-terrestrial network  is proposed  for reducing the traffic of terrestrial network. However,   spare backhaul resources is of particular importance not only for terrestrial networks but also in satellite  systems.

 To the best of the authors knowledge, the application of \acfp{LRFC} for caching content  in satellite networks  has not been proposed yet.
In this paper we study and optimize the performance of fountain codes for caching-enabled networks with satellite backhauling. We derive the average backhaul rate \footnote{we define the average backhaul rate as the average number of coded packets (output symbols) that the \acs{GEO} needs to send through the bakchaul link during the delivery phase to serve the request of a user.} for such system and optimize the cache placement. Our results show that the performance of the caching system using \acp{LRFC} is close to that of a system based on \ac{MDS} codes already for a field size $q=4$.

The rest of the paper is organized as follows. Section~\ref{sec:sysmodel} introduces the system model while in Section~\ref{sec:fountaincode} some preliminaries on \ac{LRFC} are presented. The achievable backhaul rate is presented in Section~\ref{sec:averageRate} while the optimization of the number of coded symbol to be memorized at each cache is presented in  Section~\ref{sec:placement}. In Section~\ref{sec:results} the numerical results are presented. Finally, Section~\ref{sec:Conclusions} contains the conclusions.

\section{System Model}\label{sec:sysmodel}

\begin{figure}[t]
 \includegraphics[width=0.4\textwidth]{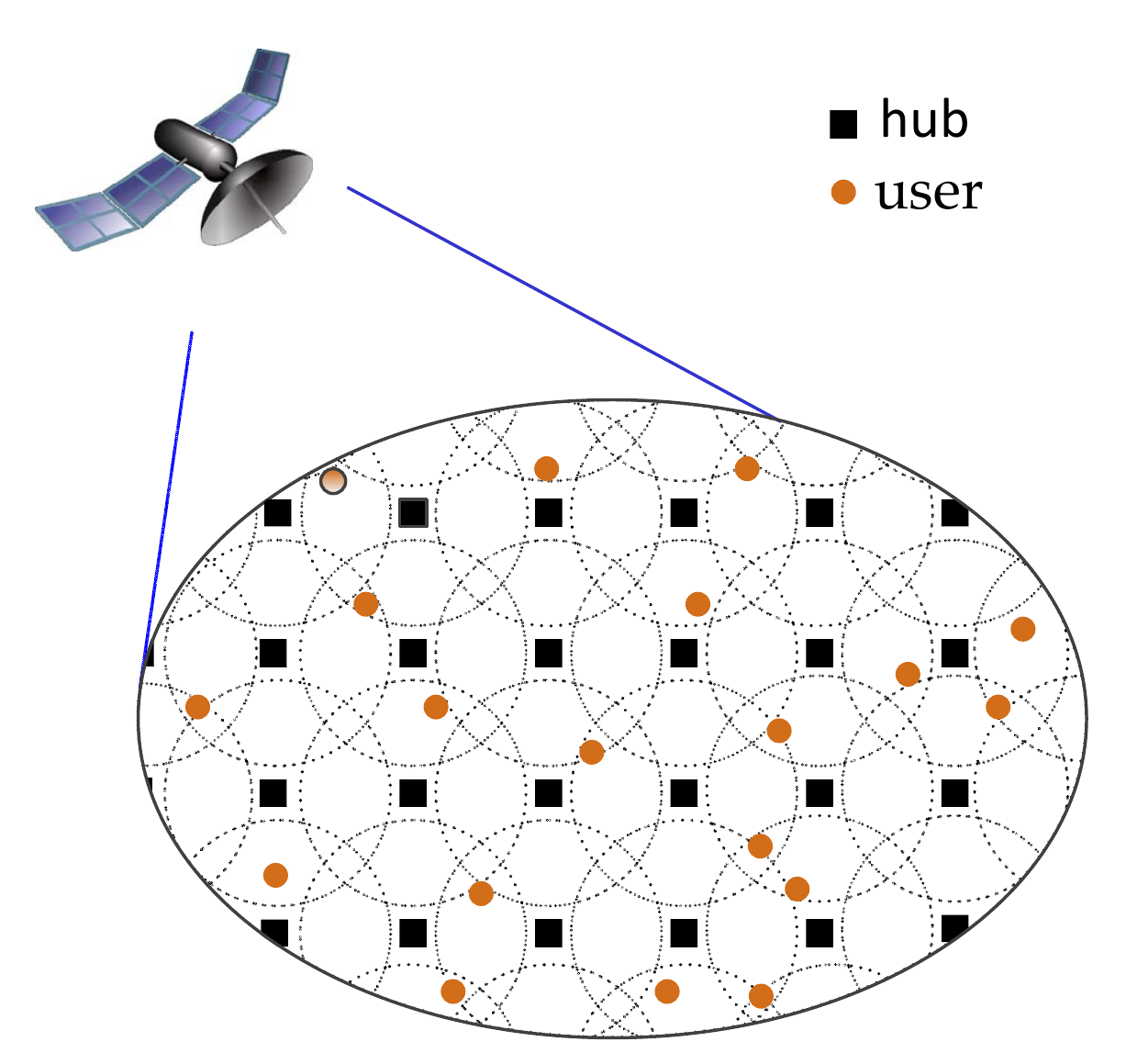}
\centering\caption{System model}
\label{fig:model}
\end{figure}

We consider a heterogeneous network composed of a single \ac{GEO} satellite and a number of hubs (e.g. terrestrial repeaters or \ac{HAPS})  with cache capabilities,  as shown in Fig.~\ref{fig:model}.
Each hub is connected to the \ac{GEO} satellite through a backhaul link.
Users are assumed to be fixed and to have a limited antenna gain so that a direct connection to the \ac{GEO} satellite is not possible. Depending on their location users may be connected to one or multiple hubs.  We denote by $\probHub{\hubs}$ the probability that a user is connected to $\hubs$ hubs.

%In particular, we consider a coverage area $\area{g}$ of the \ac{GEO}  where users are uniformly distributed. We further assume that each hub covers a circular area of radius $\radius{\hubs}$ centered around the hub.  As it is shown in Fig.~\ref{fig:model} the area of coverage of the hubs can overlap.  For simplicity, we assume that the hubs are arranged according to a uniform two dimensional grid, with spacing $\dcenter$. %We further assume that there is a hub at the center of the coverage area of the \ac{GEO}

The \ac{GEO} has access to a library of $\nfiles$ equally long files $\library = \{\libraryf{1}\, \ldots, \libraryf{\nfiles}\}$.  We assume that users request files from the library independently at random. Furthermore, we assume that the probability of file $\libraryf{j}$ being requested, $\probFile{j}$, follows a Zipf distribution with parameter $\alpha$ leading to
\[
\probFile{j} = \frac{1/j^\alpha}{\sum_i 1/i^{\alpha}}\quad \forall j = 1, ..., n.
\]
The assumption is made that all files are fragmented into $\npacket$ \emph{input symbols} (packets). The \ac{GEO} satellite encodes each file independently, using a fountain code.
Each hub has storage capability for $\memory$ files, i.e., for $k M$ packets.

We shall assume that the coded caching scheme is split into two phases.
During the placement phase the \ac{GEO} sends a number $\npcached{j}$ of output symbols from file $\libraryf{j}$ to each hub, which are cached by each hub.
Note that for the same file each hub caches  the same number of \emph{output symbols} (encoded packets), however,  the sets of output symbols cached at different hubs are different. The placement phase is assumed to be carried out offline.
During the delivery phase, a user requests a file $\libraryf{j}$  at random. In a first stage, the user downloads output symbols of $\libraryf{j}$ cached in the $\hubs$ hubs he is connected to. If the number of symbols is not enough for decoding  $\libraryf{j}$ successfully, then  the \ac{GEO} satellite generates additional output symbols. The \ac{GEO} sends the additional output symbols to the user via one of the $\hubs$ hubs he is connected to. %\sout{We define with backhaul rate the number of output symbols per file transmitted by the \ac{GEO} during the delivery phase.}
For simplicity we assume that all transmissions are error-free.

\section{Linear Random Fountain Codes}\label{sec:fountaincode}

In this work we consider the use of \acp{LRFC} for the delivery of the different files in the library. Each file is  fragmented into $k$ input   symbols,
$ (\isymb_1, \isymb_2, \hdots, \isymb_\npacket )$. For simplicity, we assume  in the following $\isymb_i \in \mathbb{F}_q$. The case of  $\isymb_i \in \mathbb{F}_q^m$, i.e. the case in which packets are m symbols long,  can be addressed as a straightforward extension. The \ac{LRFC} encoder generates a sequence of output symbols
$\mathbf{\codedpacket} = ( \codedpacket_1, \codedpacket_2, \hdots, \codedpacket_{\ell} )$,
where the number of outputs symbols $\ell$ can grow indefinitely. In particular, the $i$-th output symbols is generated as
\[
\codedpacket_{\indexi} = \sum_{a=1}^{\npacket} \coeff_{a,\indexi}   \, \isymb_{a},
\]
where the coefficients $\coeff_{a,\indexi}$ are picked  independently at random with uniform probability in $\mathbb{F}_q$.
For fixed $\ell$, \ac{LRFC} encoding can be expressed as a vector matrix multiplication
\[
\mathbf{\codedpacket} = \isymbv \G
\]
where $\isymbv$ is the vector of input symbols and $\G$ is a $k \times \ell$ matrix, whose elements $\coeff_{a,\indexi}$ are selected independently and uniformly at random in $\mathbb{F}_q$.

%\fran{The symbols produced by the \ac{LRFC} encoder are either stored in a hub during the placement phase or delivered through the backhaul link to a hub and then forwarded to a user.}
In order to download a file, a user must collect a set of $m \geq \npacket$ output symbols $\rosymbv=(\rosymb_1, \rosymb_2, \hdots, \rosymb_m)$. If we denote by $\mathcal{I} =(i_1, i_2, \hdots, i_m)$ the set of indices corresponding to the $m$ output symbols collected by the receiver we have
\[
\rosymb_r = \codedpacket_{i_r}.
\]
The user attempts decoding by solving the system of equations
\[
\rosymbv = \isymbv \Grx
\]
where $\Grx$ is a matrix corresponding to the $m$ columns of $\G$ associated to the collected output symbols i.e., the columns of $\G$ with indexes in $\mathcal{I}$. If the system of equations  admits a unique solution (i.e., if $\Grx$ is full rank), decoding is declared successful after recovering $\isymbv$, for example by means of Gaussian elimination. If $\Grx$ is rank deficient, a decoding failure is declared. In the latter case the receiver reattempts decoding after collecting one or more additional output symbols.

Let us define $\delta$ as the receiver overhead  $\delta=m-\npacket$, that is, the number of output symbols in excess to $\npacket$ that the receiver has collected.
Given $k$, $\delta$ and $q$,  the probability of decoding failure of an \ac{LRFC} is given by %\cite{Liva10:fountain}
\[
\probFail{(\npacket,\delta,q)} = 1- \prod_{\indexi=1}^{\npacket} \left(  1-\frac{q^{\indexi-1}}{q^{\npacket+\delta}}  \right)
\]
and can be tightly lower and upper bounded as  \cite{Liva10:fountain}
\begin{align}\label{eq:bounds}
\lowerb{(\delta,q)} & \leq \probFail{(\npacket,\delta,q)}  <   \upperb{(\delta,q)}
\end{align}
where
\[
\lowerb{(\delta,q)}: = q^{-\delta -1}
\]
and
\[
 \upperb{(\delta,q)}: =  \frac{1}{q-1} q^{-\delta}.
\]
Note that the bounds are independent from the number of  input symbols $\npacket$ and become tighter for increasing $q$.

 For notational convenience, in the remaining of the paper we shall use the probability of decoding success rather than the probability of decoding failure, which is simply defined as
\[
\probSuc(\npacket,\delta,q) := 1 - \probFail{(\npacket,\delta,q)}.
\]

%For calculating the average backhaul rate, we are interested on evaluating the average number of overhead $\bar{\delta}$ symbols that are needed for successfully fulfil a user request, which can be calculated as follow:
%\begin{align*}
%\bar{\delta} = & \sum_{\delta=1}^{\infty} \delta \left(1-  \probSuc{(k,\delta,q)}\right)^{\delta-1}  \probSuc{(k,\delta,q)}. \\
%\end{align*}
%
%
% In the next section, we calculate the average backhaul rate and we reformulate the optimization placement problem proposed in  \cite{bioglio:Globcom2015} to adapt to our scheme.
 \section{Average Backhaul rate} \label{sec:averageRate}

We {define the average backhaul rate as the average  number output coded symbols that the \ac{GEO} has to send during the delivery phase in order to fulfill  a user  request.

In this section, we  derive  the overhead decoding probability $\sigma_{\delta}$, that is, the probability  that a user needs exactly $k+\delta$ coded symbols to successfully decode the requested file. Then,  we  calculate the average backhaul transmission rate for a \ac{LRFC} coded caching scheme.

%Assuming that each hub has cached $\npcached{j}$ symbols of a requested file $\libraryf{j}$ then a user connected to $\hubs$ hubs collects a set of $\npcached{j} \cdot  \hubs$ different coded symbol.  As soon as  $\npcached{j} \cdot  \hubs$   are not sufficient for  successful decode the file requested then the \ac{GEO} uses the backhaul link for delivering  further coded symbols.

% In this section, we first derive the connectivity probability distribution (i.e. the probability of a random user being connected to exactly $h$ hubs), using simple geometrical considerations. Then,

\subsection{Overhead Decoding Probability} \label{susec:DecProb}

Let us denote $\sEvent_{\delta}$  the event that   the matrix $\Grx$ is full rank  when $m = k + \delta$ output symbols have been collected, where  $\Pr\{\sEvent_{\delta}\} = \probSuc{(\npacket,\delta,q)}$. Let us denote the  complementary event, i.e.   the rank of $\Grx$ is smaller than $k$,  with $\bar{\sEvent}_{\delta}$.
We are interested in deriving
\[  \sigma_{\delta} := \Pr\{\sEvent_{\delta} \mid \bar{\sEvent}_{\delta-1}\}.
\]
Starting from
\begin{align}
\Pr\{ \sEvent_{\delta}\} = &\Pr\{ \sEvent_{\delta} \mid     \sEvent_{\delta-1}\}  \Pr\{\sEvent_{\delta-1}\} \\
+&   \Pr\{\sEvent_{\delta}\mid \bar{\sEvent}_{\delta-1}\} \Pr\{\bar{\sEvent}_{\delta-1}\} \\
= &  \Pr\{\sEvent_{\delta-1}\}+
  \Pr\{\sEvent_{\delta}\mid \bar{\sEvent}_{\delta-1}\} \Pr\{\bar{\sEvent}_{\delta-1}\} \\
\end{align}
we have that
\begin{align}
\Pr\{\sEvent_{\delta}\mid \bar{\sEvent}_{\delta-1}\}
& =   \frac{\Pr\{ \sEvent_{\delta}\}  - \Pr\{\sEvent_{\delta-1}\}}{\Pr\{\bar{\sEvent}_{\delta-1}\}} \\
& = \frac{1 - \Pr\{\bar{\sEvent}_{\delta}\} - (1 - \Pr\{\bar{\sEvent}_{\delta-1}\} )}  {\Pr\{\bar{\sEvent}_{\delta-1}\}}\\
& =   1 - \frac{\Pr\{\bar{\sEvent}_\delta\}}{\Pr\{\bar{\sEvent}_{\delta-1}\}}
\end{align}
which can be rewritten as
\begin{equation}
\sigma_{\delta} = 1 - \frac{1- \probSuc{(\npacket,\delta,q)}}{1- \probSuc{(\npacket,\delta-1,q)}}.\label{eq:Ps}
\end{equation}
 The expression in \eqref{eq:Ps} holds for $\delta \geq 0$, whereas for $\delta < 0$ we have $\sigma_{\delta}=0$.
\newline
Bounds on equation~\eqref{eq:Ps} can be obtained from~\eqref{eq:bounds}. In particular, for $\delta \geq 0$  we can write
\begin{equation}
1-\frac{  \upperb{(\delta,q)}}{ \lowerb{(\delta-1,q)}}< \sigma_{\delta} <  1 - \frac{  \lowerb{(\delta,q)}}{ \upperb{(\delta-1,q)}}
\end{equation}
yielding
\begin{equation}\label{eq:sigmabounds}
1-\frac{1}{q-1}< \sigma_{\delta} <  1 -  \frac{q-1}{q^2}.
\end{equation}
Note that the bounds are independent of the overhead ${\delta}$ (for non negative $\delta$) and  become tight as $q$ grows. Note also that for $q=2$ the lower bound becomes $0$ and, hence, it loses significance.
\subsection{Overhead Average}
 Let us denote as  $\Delta$ the random variable associated   to the  average  number of symbols in excess to $k$  that a user needs to recover the requested content and  let us also denote as $\delta$ its realization. We can calculate  the average overhead as follows
\begin{align}\label{eq:delta_av}
\Expd{} & = \sum_{\delta=0}^{\infty} \delta \left[ \prod_{i=0}^{\delta-1}  \left(1-\sigma_{i}\right) \right]\sigma_{\delta}.\\
\end{align}

By using \eqref{eq:sigmabounds} in \eqref{eq:delta_av}, $\Expd{}$ can be upper bounded as
\begin{align}
\Expd{} < \frac{q-1}{(q-2)^2} \left(1-\frac{q-1}{q^2}\right):= \delta_u.
\label{eq:delta_u}
\end{align}

%If we denote the lower and upper  bounds of~\eqref{eq:sigmabounds} respectively as
%\[\bound{l} = 1-\frac{1}{q-1}\]
%and
%\[\bound{u} = 1-\frac{q-1}{q^2}\]
%we can define
%\begin{align}\Expd{u} = \sum_{\delta=0}^{\infty} \delta \left(1-\bound{l}\right)^{\delta}  \bound{u}
%\label{eq:delta_u}
%\end{align}
%where
%\begin{align}
%\Expd{} & \leq \Expd{u}  \\
%& = {\bound{u}} \sum_{\delta=0}^{\infty} \delta \left(1-\bound{l}\right)^{\delta}  \\
%& = {\bound{u}} \frac{1-\bound{l}}{\bound{l}^2}  \\
%& = \frac{q-1}{(q-2)^2} \left(1-\frac{q-1}{q^2}\right) \label{eq:delta_uleq}
%\end{align}
%where $\Expd{u}$ decreases as $q$ grows.
%%
\subsection{Backhaul Rate}
Let $Z$ be the random variable associated to the number of output symbols for a file requested by a user in the hubs he is connected to, and let $z$ be its realization.  Let $H$ be the random variable associated to the number of hubs a user is connected to, being $h$ its realization. Finally, let $J$ be the random variable associated to the index of the file requested by a user, being $j$ its realization. We have
\begin{align}\label{eq:Z}
  P_{Z|J,H}(z | j, h) = \begin{cases}
                              1  & \mbox{if } z= \npcached{j} \, h\\
                              0  & \mbox{otherwise}
                            \end{cases}
\end{align}
where we recall that  $\npcached{j}$ stands for the number of coded symbols from file $j$ stored in every hub. The probability mass function of $Z$ is
\begin{align}\label{eq:z_2}
    P_Z(z)    &= \sum_j \sum_h  P_{Z|J,H}(z | j, h) P_J(j)  P_H(h)  \\
   & =  \sum_j \sum_h \probFile{j} \, \probHub{\hubs} \,  P_{Z|J,H}(z | j, h).
\end{align}
We are interested in deriving the distribution of the backhaul rate, i.e, the number of output symbols which have to be sent over the backhaul link to serve the request of a user, which we denote by random variable $T$. If we condition $T$ to $Z$, it is easy to see how the probability of $T=t$ corresponds to the probability that decoding succeeds when the user has received exactly $t$ output symbols from the backhaul  link in excess to the  $z$ output symbols it received from the hubs through local links, that is, when $m=z+t$.
\newline
 In order to derive $P_{T|Z}(t|z)$ we shall distinguish two cases.
\newline
 If $z \leq k$, then
\begin{align}\label{eq:t_cases1}
&  P_{T|Z}(t|z) =
 \begin{cases}
 \displaystyle     \prod_{i=0}^{z-k+t-1}(1-\sigma_i) \sigma_{z-k+t} & \mbox{if } t \geq 0 \\
                 0  & \mbox{otherwise}.
                        \end{cases}
\end{align}
If $z>k$, then
\begin{align}\label{eq:t_cases2}
&  P_{T|Z}(t|z) =
\begin{cases}
   \displaystyle  \sum_{j=0}^{z-k} \Bigg(    \displaystyle    \prod^{j -1}_{i=0}(1-\sigma_i) \sigma_{j} \Bigg)  & \mbox{if } t = 0 \\
    \displaystyle  \prod^{z-k+t-1}_{i=0}(1-\sigma_i) \sigma_{z-k+t}   & \mbox{if t} > 0 \\
                   0 & \mbox{otherwise}.
  \end{cases}
\end{align}

We are after the expectation of  $T$, which is obtained as

\begin{align}\label{eq:t}
  \mathbb{E}[T] &=  \sum_t t P(t) \\
   &= \sum_t t \sum_{z=0}^{\infty} \Pro_{T|Z}(t|z) \Pro_Z(z) \\
   &= \sum_t t \Bigg( \sum_{z=0}^{k} \Pro_{T|Z}(t|z) \Pro_Z(z)+ \mkern-6mu \sum_{z=k+1}^{\infty} \Pro_{T|Z}(t|z) \Pro_Z(z)  \Bigg)  \\
\end{align}
where in the last equality we distinguished two different cases, $z\leq k$  and  $z>k$.
Let us define $\bar{T}_1$ and $\bar{T}_2$ as
\begin{align}
  \bar{T}_1 :&=  \sum_t t  \sum_{z=0}^{k}\Pro_{T|Z}(t|z) \Pro_Z(z) \\
  \bar{T}_2 :&=  \sum_t t \sum_{z=k+1}^{\infty} \Pro_{T|Z}(t|z) \Pro_Z(z)
\end{align}
so that
\begin{align} \label{eq:T1T2}
 \mathbb{E}[T] = \bar{T}_1 + \bar{T}_2.
\end{align}

If we introduce the variable change $\delta = z-k+t$ in the expression of $\bar{T}_1$, we obtain
\begin{align}\label{eq:a1}
  \bar{T}_1 & =
  \sum_{z=0}^{k}  \Pro_Z(z)  \sum_{\delta=z-k}^{\infty} (\delta -z+k )\prod_{i=0}^{\delta-1}(1-\sigma_i) \sigma_{\delta}   \\
  & \stackrel{(\mathrm{a})}{=}  \sum_{z=0}^{k} \Pro_Z(z)  \sum_{\delta=0}^{\infty} (\delta -z+k )\prod_{i=0}^{\delta-1}(1-\sigma_i) \sigma_{\delta} \\
  &=    \sum_{z=0}^{k}  \Pro_Z(z) \Bigg( \sum_{\delta=0}^{\infty} \delta \prod_{i=0}^{\delta-1}(1-\sigma_i) \sigma_{\delta} \\
 &  +  (k-z) \sum_{\delta=0}^{\infty}  \prod_{i=0}^{\delta-1}(1-\sigma_i) \sigma_{\delta} \Bigg)  \\
  &\stackrel{(\mathrm{b})}{=}    \sum_{z=0}^{k} \Pro_Z(z)   (  \Expd{} +  k-z )   \\
  &= ( \Expd{} +  k)  \Pr\{Z \leq k\}   -  \sum_{z=0}^{k} z   \Pro_Z(z)  \label{eq:T1}
\end{align}
where equality $(\mathrm{a})$ is due to $\sigma_{\delta} = 0$ for $\delta < 0$ and equality $(\mathrm{b})$ is due to
\[
\sum_{\delta=0}^{\infty}  \prod_{i=0}^{\delta-1}(1-\sigma_i) \sigma_{\delta} = 1.
\]
Introducing the same variable change in the expresion of $\bar{T}_2$ we have
\begin{align}\label{eq:a2}
  \bar{T}_2 &=   \sum_{z=k+1}^{\infty}  \Pro_Z(z)  \sum_{\delta=z-k}^{\infty} (\delta -z+k )\prod_{i=0}^{\delta-1}(1-\sigma_i) \sigma_{\delta} \\
  %& = \sum_{z=k+1}^{\infty}  \Pro_Z(z)  \sum_{\delta=z-k}^{\infty} (\delta -z+k )\prod_{i=0}^{\delta-1}(1-\sigma_i) \sigma_{\delta} \\
  & = \sum_{z=k+1}^{\infty}  \Pro_Z(z)  \Bigg( \sum_{\delta=z-k}^{\infty}  \delta \prod_{i=0}^{\delta-1}(1-\sigma_i) \sigma_{\delta} \\
  & + \sum_{\delta=z-k}^{\infty} (k-z) \prod_{i=0}^{\delta-1}(1-\sigma_i) \sigma_{\delta} \Bigg)\\
  & \leq \sum_{z=k+1}^{\infty}     \Pro_Z(z)   \Expd{}  =  \Expd{} \Pr\{Z > k\}
  \label{eq:upT}
\end{align}
where the inequality is due to
\[
 \sum_{\delta=z-k}^{\infty}  \delta \prod_{i=0}^{\delta-1}(1-\sigma_i) \sigma_{\delta} \leq \Expd{}
\]
and  to
\[
\sum_{\delta=z-k}^{\infty} (k-z) \prod_{i=0}^{\delta-1}(1-\sigma_i) \sigma_{\delta} \leq 0
\]
since $z>k$.
\newline
The expression of the average backhaul rate given by
\begin{align}
 \mathbb{E}[T] &=  ( \Expd{} +  k)  \Pr\{Z \leq k\}   -  \sum_{z=0}^{k} z   \Pro_Z(z) \\ & +   \sum_{z=k+1}^{\infty}  \Pro_Z(z)  \Bigg( \sum_{\delta=z-k}^{\infty}  \delta \prod_{i=0}^{\delta-1}(1-\sigma_i) \sigma_{\delta}
\\ & +  \sum_{\delta=z-k}^{\infty} (k-z) \prod_{i=0}^{\delta-1}(1-\sigma_i) \sigma_{\delta} \Bigg) \label{eq:ET}.
\end{align}

Finally, we can upper bound the average backhaul rate by use of \eqref{eq:delta_u}, \eqref{eq:T1} and \eqref{eq:upT} in \eqref{eq:T1T2} as
 \begin{align}
 \mathbb{E}[T]  \leq  \delta_u + k  \Pr\{Z \leq k\} -  \sum_{z=0}^{k} z   \Pro_Z(z)  \label{eq:upTfinal}.
 \end{align}

\section{  \ac{LRFC}  Placement Optimization}\label{sec:placement}

The \ac{LRFC} placement   problem calls for
minimizing the average backhaul rate during the delivery phase. To this end, we want to optimize the number of coded symbols per file that each hub has to cache.  We present in this section the placement optimization problem adapted to a \ac{LRFC} cached scheme based on the optimization problem proposed for \ac{MDS} codes in~\cite{bioglio:Globcom2015}.

The optimization problem can be written as
\begin{align}\label{eq:optProblem}
\underset{\npcached{1}, \dots, \npcached{\nfiles}}  \min &  \quad  \mathbb{E}[T]  \\
 \text{s.t. }& \quad \sum_{j=1}^{\nfiles}  \npcached{j}    = \memory \npacket,  \quad     \npcached{j} \in  \mathbb{N}_0.
\end{align}
The first constraint specifies that the total number of stored coded symbols should be equal to the size cache. The second constraints accounts for the discrete nature of the optimization variable.

Solving exactly the optimization problem requires evaluating \eqref{eq:t}, which can be computationally complex when $k$ and $z$ are large. Hence, as an alternative to minimizing the average backhaul rate, we propose minimizing its upper bound  in~\eqref{eq:upTfinal}, which leads to the following optimization problem
\begin{align}
\underset{\npcached{1}, \dots, \npcached{\nfiles}}  \min &  \Bigg[   \delta_u + k \Pr\{Z \leq k\} - \sum_{z=0}^{k} z  \Pro_Z(z)  \Bigg] \\
 \text{s.t. }& \quad \sum_{j=1}^{\nfiles}  \npcached{j}    = \memory \npacket  \quad  \npcached{j} \in \mathbb{N}_0 . \label{eq:optProblem2}
\end{align}
Since the upper bound on $\mathbb{E}[T]$  in \eqref{eq:upTfinal} relies on the upper and lower bounds in \eqref{eq:bounds}, which are tight, we expect the result of the optimization problem in \eqref{eq:optProblem2} to be close to the result of the optimization problem in \eqref{eq:optProblem}. 
\section{Results}\label{sec:results}
\begin{figure}[t]
 \includegraphics[width=0.48\textwidth]{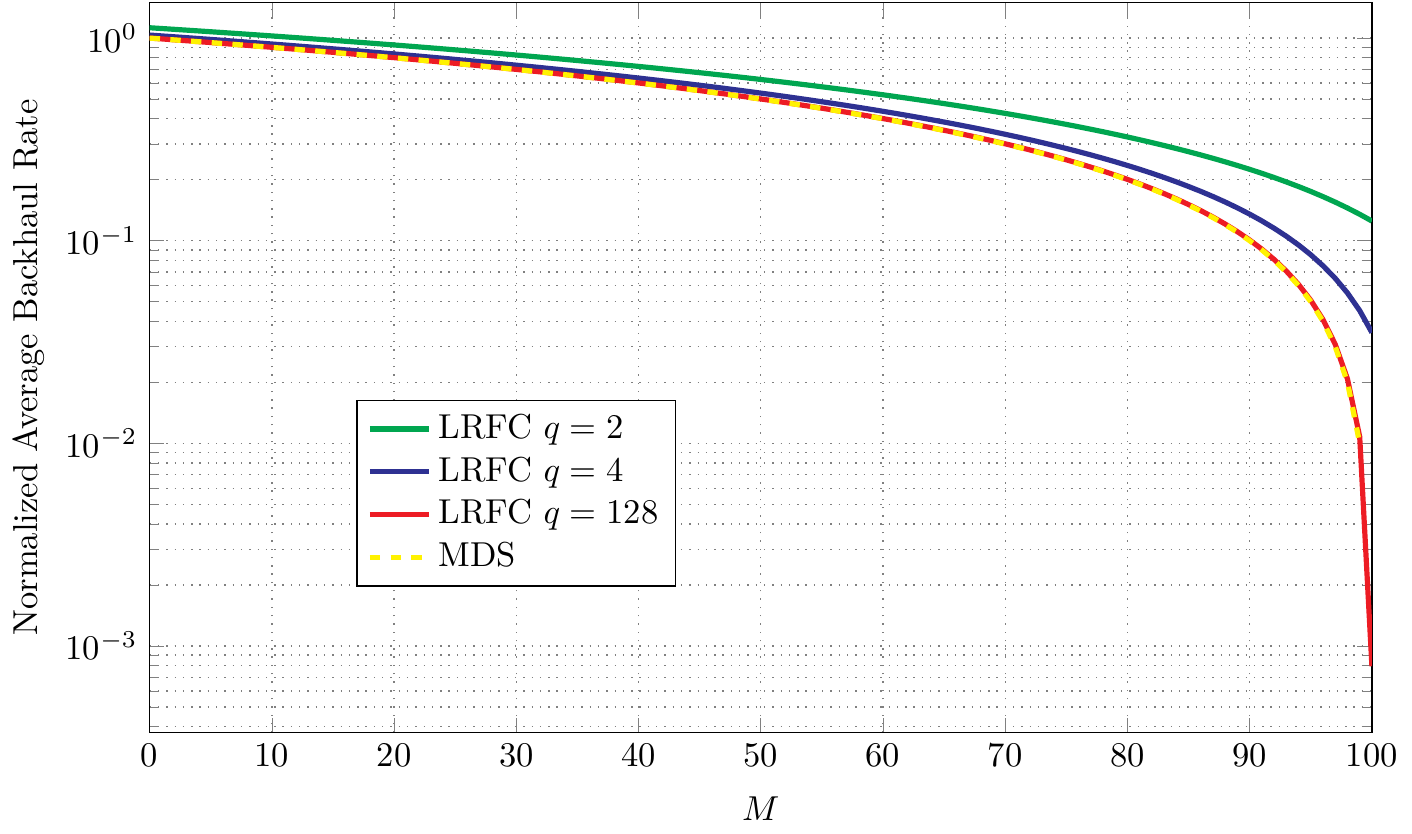}
\centering \caption{Normalized average backhaul rate as a function of memory size $M$ for  \ac{MDS} and \ac{LRFC} codes over $\mathbb{F}_q$ for $q=2,4,128$ given  $n=100$, $k=10$, $\alpha=0$ and $\gamma_1=1$. Solid curves represent \ac{LRFC} schemes while the dashed curve represents the \ac{MDS} scheme.}
\label{fig:R_u_1}
\end{figure}

In this section, we numerically evaluate the normalized average backhaul rate, which we define as $\mathbb{E}[T]/k$.

In all the setups, we consider that users are uniformly distributed within the coverage area of the satellite and border effects are neglected.
 We consider  that each hub covers a circular area of radius $\radius{}$ centered around the hub.   For simplicity, we assume that the hubs are arranged according to a uniform two dimensional grid, with spacing $\dcenter$.  Unless otherwise specified, we assume $\radius{}=60$ km and   $\dcenter=45$ km. Thus, the coverage areas of different hubs partially overlap, as it can be observed in Fig.~\ref{fig:model}. After some simple geometrical calculations we obtain that the following connectivity distribution
\begin{align}\label{eq:gammadis}
& \gamma_1 = 0.2907,  \,  \gamma_2 =0.6591, \, \\
& \gamma_3 = 0.0430, \, \gamma_4 =0.0072.
\end{align}

We first evaluate the tightness of the upper bound \eqref{eq:delta_u} on average overhead. Table~\ref{tab:tabdelta} shows  $\Expd{}$  for different values of $q$  are shown. The values in the second column were numerically derived from equation~\eqref{eq:delta_av} while values in the third column were derived from the bound in equation~\eqref{eq:delta_u}. We can see that the bound becomes tighter for increasing $q$.

\begin{center}
\begin{table}[t] \centering
\caption{Average overhead required for successful decoding for a \ac{LRFC} with $k = 10$}\label{tab:tabdelta}
\begin{tabular}{ccc}
\hline
  $q$   &  $\Expd{}$  & upper bound \eqref{eq:delta_u}\\
\hline
$2$   & 1.1981   &       - \\
$4 $  & 0.3490   &   0.6094  \\
$8 $  & 0.1447   &   0.1792  \\
$16$  & 0.0669   &   0.0720  \\
$32$  & 0.0323   &   0.0334  \\
$64$  & 0.0159   &   0.0161  \\
$128$ & 0.0079   &   0.0079  \\
\hline
\end{tabular}
\end{table}
\end{center}

% \subsection{Average Backhaul Rate}
In the first scenario, we study how  the cache size $M$ impacts in the average backhaul rate.  In this setup, we assume all user are connected to exactly one hub, i.e.  $\gamma_1=1$, furthermore, we  consider that file popularity is uniformly  distributed (i.e Zipf distributed with parameter $\alpha =0$). Moreover, we assume a library size $n=100$ and we assume that each file is fragmented into $k=10$ input symbols.
We optimized  the number of LRFC coded symbols $\npcached{j}$  cached  at each hub by solving the problem \eqref{eq:optProblem2} for  $q=2$, $q=4$ and $q=128$ and we calculated the average backhaul rate the fountain coding caching scheme. As a benchmark we used the \ac{MDS} caching scheme from~\cite{bioglio:Globcom2015}.

In Fig.~\ref{fig:R_u_1} the normalized average backhaul  rate is shown as a function of the memory size $M$. We can observe how the penalty on the average  rate for using \ac{LRFC} with respect to a \ac{MDS} code becomes smaller for increasing $q$ and already for $q = 128$ is almost negligible. We remark that for $M=100$ the cache size coincides with the library size, hence, the backhaul  rate for the \ac{MDS} scheme becomes zero, whereas for the \ac{LRFC} schemes the average backhaul  rate coincides with the average overhead.

In our second setup, we assume that users can be connected to multiple hubs. We consider the connectivity distribution  given in~\eqref{eq:gammadis} and file popularity  distribution with parameter $\alpha =0.8$. The library size is set to $n=100$ and the number of input symbols is $k=10$, the same as in the previous scenario. The optimal cache placing is computed for each \ac{LRFC} scheme as well as for the \ac{MDS}. %Given $\npcached{j}$  we evaluate the average backhaul rate.

In Fig.~\ref{fig:R_M} we show the impact of memory size $M$  on the normalized average backhaul  rate  for different code caching schemes when users can be connected to multiple hubs.
Similarly to the previous scenario,   we see that for sufficiently large $q$ the performance of the \ac{LRFC} caching scheme approaches that of the MDS code. Note that since a MDS code achieves the best possible performance, this result shows implicitly that solving the optimization problem in \eqref{eq:optProblem2} yields a solution that is close to that of solving the  optimizaiton problem in \eqref{eq:optProblem}. We further observe that  \ac{LRFC}  caching     with storage capabilities  equal  to 10\% of the library size can reduce the average backhaul  rate for at least 40\%  with respect to a system with no caching ($M=0$).

\begin{figure}[t]
 \includegraphics[width=0.48\textwidth]{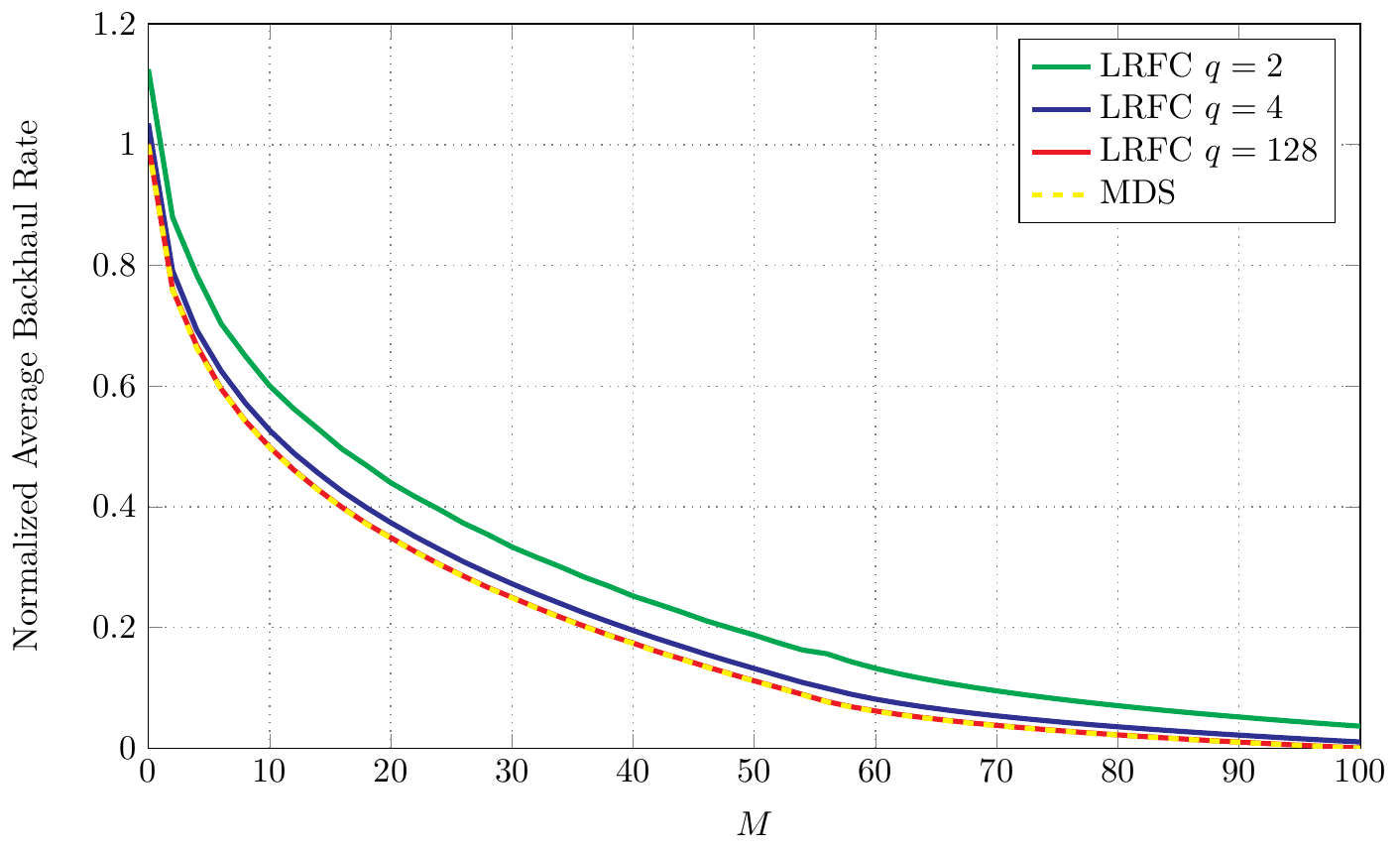}
\centering \caption{Normalized average backhaul  rate as a function of memory size $M$ for \ac{MDS} and \ac{LRFC} codesover $\mathbb{F}_q$ for $q=2,4,128$ given $n=100$, $k=10$, $\alpha=0.8$ and $\gamma_1=0.2907$, $\gamma_2=0.6591$, $\gamma_3=0.0430$, $\gamma_4=0.0072$.}
\label{fig:R_M}
\end{figure}

For the same connectivity distribution,   a fixed memory size $M=10$ and library size $n=100$, we  investigate how the file distribution impacts on the average backhaul  rate.
In Fig.~\ref{fig:R_alpha_theta} the normalized average backhaul  rate is shown as a function of the file parameter distribution $\alpha$. As expected, when $\alpha$ increases caching schemes become more efficient since the majority of the requests is concentrated in a small number of files.
Looking at the figure we can observe how for $\alpha=0.2$, a \ac{LRFC} in  $\mathbb{F}_2$ requires roughly 12\% more transmissions over backhaul link   than  a  \ac{LRFC} in  $\mathbb{F}_{128}$. For  $\alpha=1.5$ the \ac{LRFC} of order $q=2$ requires only 4.7\% more than in $q=128$.
\begin{figure}[t]
 \includegraphics[width=0.48\textwidth]{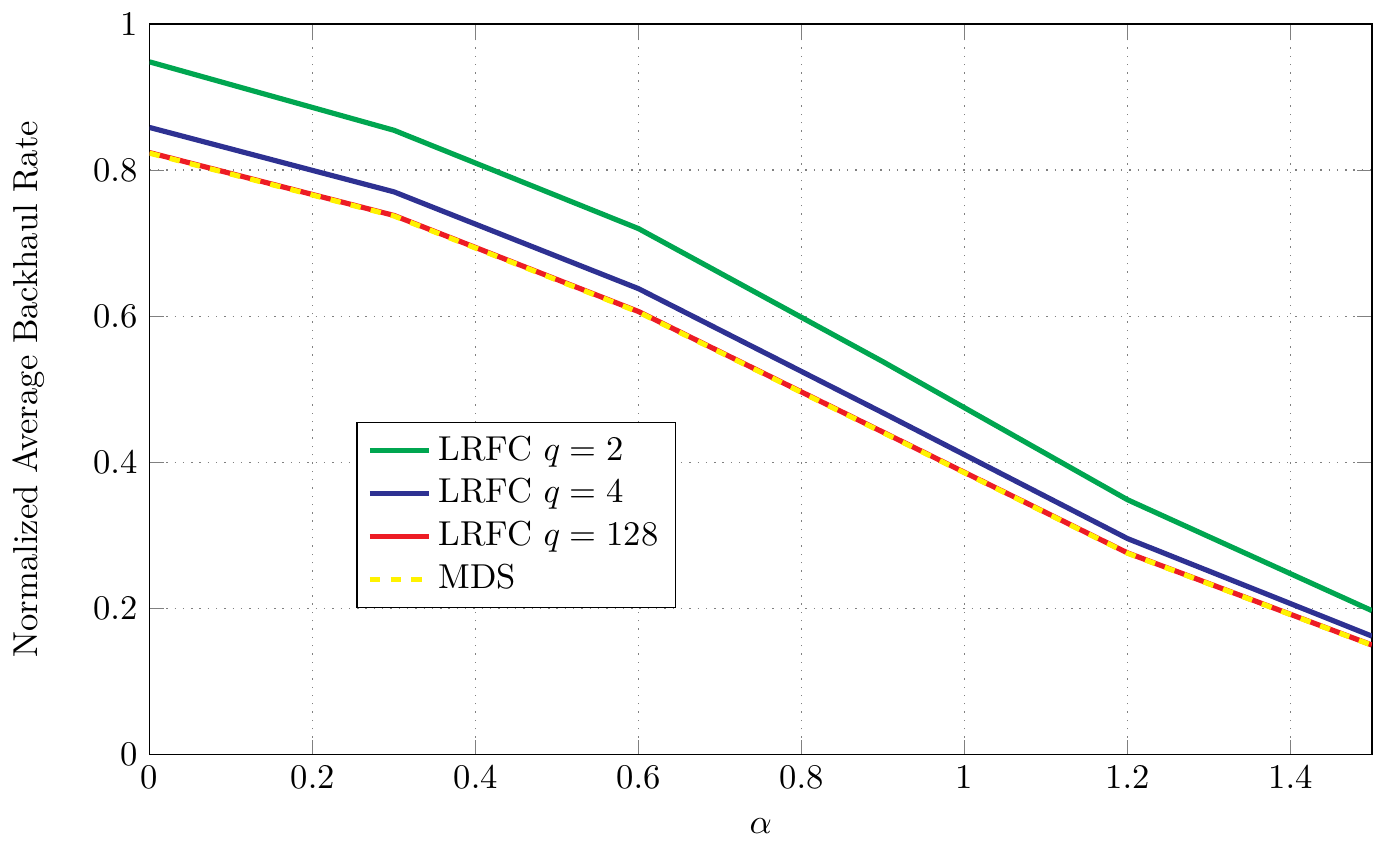}
\centering \caption{Normalized average backhaul  rate as a function of the file parameter distribution $\alpha$ for \ac{MDS} and \ac{LRFC} codes over $\mathbb{F}_q$ for $q=2,4,128$ given $n=100$, $k=10$, $M=10$ and $\gamma_1=0.2907$, $\gamma_2=0.6591$, $\gamma_3=0.0430$, $\gamma_4=0.0072$.}
\label{fig:R_alpha_theta}
\end{figure}

In our last setup we consider $\alpha=0.8$, $M=10$, $k=10$ and the distribution given in~(\ref{eq:gammadis}). We evaluate the average backhaul  rate   for different cardinalities of the library.
In Fig.~\ref{fig:R_n} the nomalized average backhaul  rate is shown as a function of the library size.
For a fixed memory size the average backhaul  rate increases as the library size increases.  As it can be observed, also in this case the proposed \ac{LRFC} caching scheme  performs similarly to a \ac{MDS} scheme.
\begin{figure}[t]
 \includegraphics[width=0.48\textwidth]{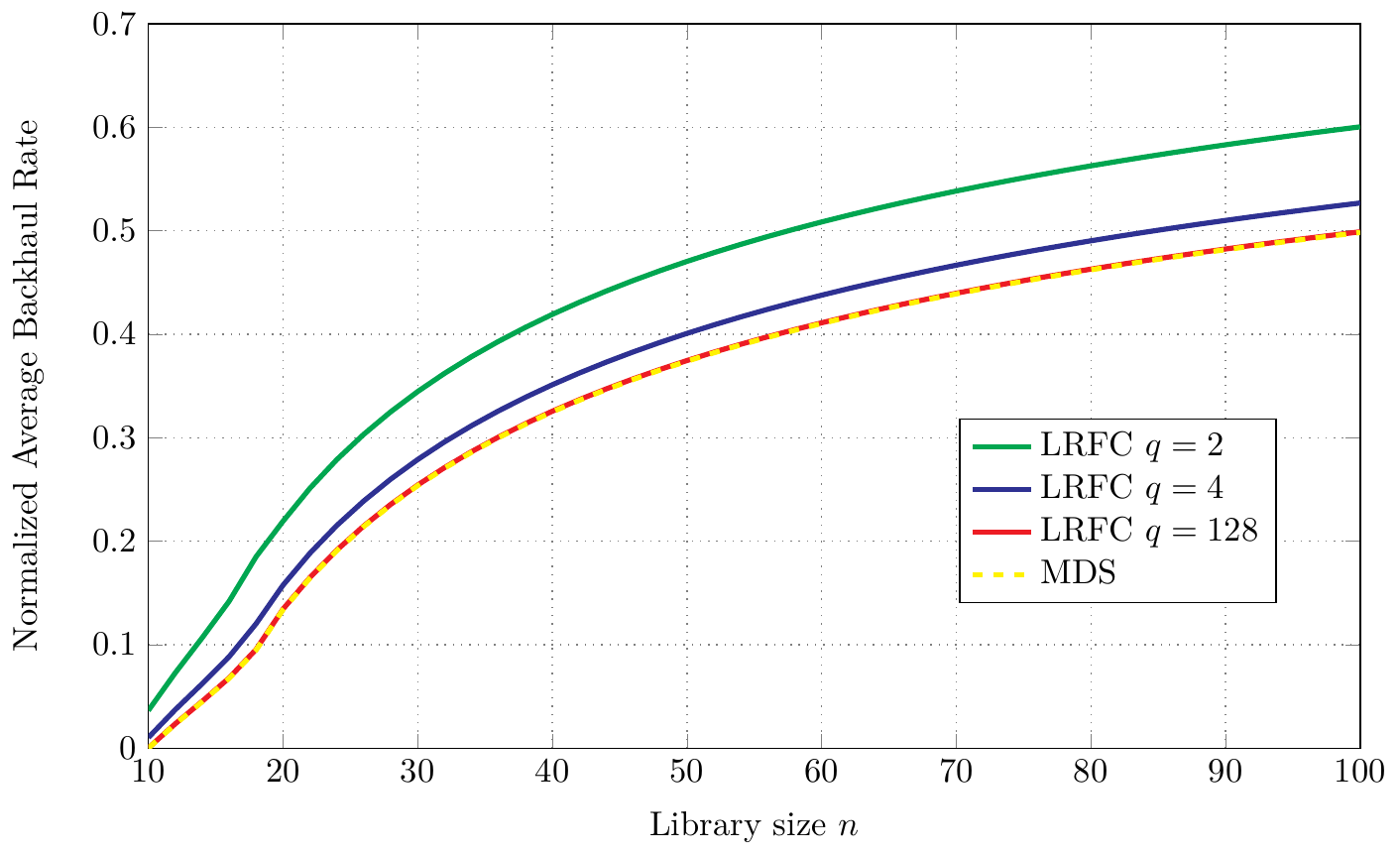}
\centering \caption{Normalized average backhaul  rate as a function of the library size $n$ for \ac{MDS} and \ac{LRFC} codes over $\mathbb{F}_q$ for $q=2,4,128$ given $\alpha=0.8$, $k=10$, $M=10$ and $\gamma_1=0.2907$, $\gamma_2=0.6591$, $\gamma_3=0.0430$, $\gamma_4=0.0072$.}
\label{fig:R_n}
\end{figure}

\section{Conclusions}\label{sec:Conclusions}
In this work we analyzed fountain code schemes for caching content at the edge. We considered a heterogeneous satellite network, composed of a GEO satellite and a number of hubs which can cache content.  We focus on reducing the average rate in the backhaul link which connects the GEO to the hubs. For this setting, we derived the analytical expression of the average backhaul rate as well as an upper bound of it.  Making use of this upper bound, we formulated the cache placement optimization problem for a fountain coding caching scheme using \acfp{LRFC}. Finally, we presented simulation results where we  compared  the performance of the  \ac{LRFC} scheme with a \ac{MDS} scheme. Our simulation results indicate that the performance of the \ac{LRFC} caching scheme built over a finite field moderate order  approaches that of the \ac{MDS} caching scheme.
%Our interest on applying \ac{LRFC} scheme instead of a \ac{MDS} is justified  by the fact that \change{ if we consider losses due to the variation of the channel then a  \ac{LRFC} can always generate new coded output symbols. }

%\section*{Acknowledgement}
%DAAD Scholarship

\bibliographystyle{IEEEtran}
\bibliography{IEEEabrv,references}

%\atColsEnd{\vskip-1pt}

%\hspace{1cm}
\flushend

\end{document}